\documentclass[sn-aps]{sn-jnl}

\jyear{2022}%

\theoremstyle{thmstyleone}%
\theoremstyle{thmstyletwo}%

\theoremstyle{thmstylethree}%

\usepackage{csquotes} 
\usepackage{hyperref}
\usepackage{amssymb, amsmath, mathptmx} 
\usepackage[version=4]{mhchem}  
\usepackage{gensymb}  
\usepackage{siunitx}
\usepackage{placeins} 

\usepackage{setspace}
\makeatletter 
\renewcommand\@biblabel[1]{#1.} 
\makeatother

\newcommand{\FGT}{\ce{Fe_3GeTe_2}}

\begin{document}

\title{Field-Dependent Magnetic Domain Behavior in van der Waals \ce{Fe_3GeTe_2}}


\author[1]{\fnm{Yue} \sur{Li}}

\author[2]{\fnm{Rabindra} \sur{Basnet}}

\author[3]{\fnm{Krishna} \sur{Pandey}}

\author[2,3]{\fnm{Jin} \sur{Hu}}

\author[4]{\fnm{Wei} \sur{Wang}}

\author[4]{\fnm{Xuedan} \sur{Ma}}

\author[1,5]{\fnm{Arthur R. C.} \sur{McCray}}

\author[1,6]{\fnm{Amanda K.} \sur{Petford-Long }}

\author[1,6,7]{\fnm{Charudatta} \sur{Phatak}}

\affil[1]{\orgdiv{Materials Science Division}, \orgname{Argonne National Laboratory}, \orgaddress{\city{Lemont}, \postcode{60439}, \state{IL}, \country{USA}}}

\affil[2]{\orgdiv{Department of Physics}, \orgname{University of Arkansas}, \orgaddress{ \city{Fayetteville}, \postcode{72701}, \state{AR}, \country{USA}}}

\affil[3]{\orgdiv{Materials Science and Engineering Program}, \orgname{University of Arkansas}, \orgaddress{ \city{Fayetteville}, \postcode{72701}, \state{AR}, \country{USA}}}

\affil[4]{\orgdiv{Center for Nanoscale Materials}, \orgname{Argonne National Laboratory}, \orgaddress{ \city{ Lemont}, \postcode{60439}, \state{IL}, \country{USA}}}

\affil[5]{\orgdiv{Applied Physics Program}, \orgname{Northwestern University}, \orgaddress{ \city{Evanston}, \postcode{60208}, \state{IL}, \country{USA}}}

\affil[6]{\orgdiv{Department of Materials Science and Engineering}, \orgname{Northwestern University}, \orgaddress{ \city{Evanston}, \postcode{60208}, \state{IL}, \country{USA}}}

\affil[7]{e-mail: cd@anl.gov}


\abstract{Two-dimensional magnetic van der Waals (vdW) materials can show a variety of topological nontrivial spin textures, such as Bloch- or N\'eel-type stripe, skyrmion or bubble domains under certain external stimuli. It is critical to understand the magnetic domain behavior in vdW materials in order to control their size, and density in response to external stimuli such as electric and magnetic fields. Here we examine the magnetic field dependence of topologically non-trivial magnetization spin textures in vdW \ce{Fe_3GeTe_2}. N\'eel-type stripe domains and skyrmions are formed depending on the magnetic field-cooling protocol used during $in$-$situ$ Lorentz transmission electron microscopy (LTEM) experiments. Use of quantitative reconstruction of magnetic induction maps, and micromagnetic simulations, allow for understanding the LTEM results of N\'eel-type stripe domains as well as skyrmions.  In addition, the deformation of skyrmion contrast is observed as a result of the introduction of an in-plane magnetic field. We demonstrate the stability of the stripe domains and skyrmions in response to externally applied magnetic field due to energy barrier for domain wall annihilation. Our results establish an understanding of the energy landscape that governs the behavior of the topologically non-trivial spin textures in vdW materials which can be harnessed for spintronic applications.}


\maketitle

\section*{INTRODUCTION}

Two-dimensional (2D) magnetic van der Waals (vdW) materials are the magnetic counterparts of more well-known 2D materials\cite{novoselov20162d, geim2013van} such as \ce{NbSe_2}\cite{efetov2016specular} and \ce{MoS_2}\cite{splendiani2010emerging}. Due to their strong intrinsic spin fluctuations and atomically-layered structure, there are exciting opportunities to explore fundamental magnetic physics in low dimensions\cite{liu2020vapor, carteaux1995crystallographic, liu2019anisotropic, huang2017layer, zhuang2016strong, leon2016magnetic, han2019topological}  and also potential application for spintronic device development\cite{xia2017recent, dankert2017electrical}. \ce{Fe_3GeTe_2} has recently emerged as a promising magnetic vdW material. It belongs to the P6$_3$/mmc space group and has a hexagonal crystal structure in the $ab$ plane and Te-\ce{Fe_3}Te-Te sandwich layers along the $c$ axis\cite{kim2018large, park2021neel, ding2019observation, deiseroth2006fe3gete2}. As an iron-rich layered compound, \ce{Fe_3GeTe_2} has been reported to possess itinerant ferromagnetism with strong perpendicular anisotropy and a Curie temperature that varies between 100 - 400 K depending on sample thickness or strain\cite{fei2018two, kim2018large, ding2019observation, nguyen2018visualization, zhuang2016strong,leon2016magnetic, chen2013magnetic, kim2019antiferromagnetic}. \ce{Fe_3GeTe_2} is being considered as a potential candidate for spintronic devices\cite{kim2018large, park2021neel, pei2020anomalous,johansen2019current, wang2019current} and also as an electrocatalyst\cite{rezaie2021abundant}. Thus, it is necessary to understand the local magnetization behavior in \ce{Fe_3GeTe_2} in response to external stimuli, such as strain, an external magnetic field, applied current and temperature, in order to gain insight into the energy landscape and the governing energy terms that control its magnetization behavior.  

There have been several previous investigations of direct visualization of N\'eel- or Bloch- type magnetic domains arising from intrinsic interactions or interfacial interactions in \ce{Fe_3GeTe_2} under various external excitations \cite{ding2019observation,pei2020anomalous,wu2020neel,yang2020creation, Peng2021Tuable, park2021neel}. In this work, we report on the field-dependent domain behavior of $\FGT$ explored using $in$-$situ$ cryo Lorentz transmission electron microscopy (LTEM). Using quantitative retrieval of magnetic induction maps, in combination with micromagnetic simulations, we show that the resulting magnetization configuration is of N\'eel-type and shows stripe domains or skyrmion lattices depending on the magnetic field cooling through the transition temperature. We also observe the deformation of skyrmions resulting from a small in-plane bias field introduced in the experiment. Apart for the formation of stripe domains and skyrmions, we further study how these domains respond to an externally applied magnetic field at a low temperature in deep ferromagnetic regime. Our results demonstrate the control of magnetic spin textures by manipulating the applied field.

\section*{RESULTS AND DISCUSSIONS}

\subsection*{N\'eel-type magnetic domains}\label{subsec:Neel}

\begin{figure}[!htbp]%
\centering
\includegraphics[width=1\textwidth]{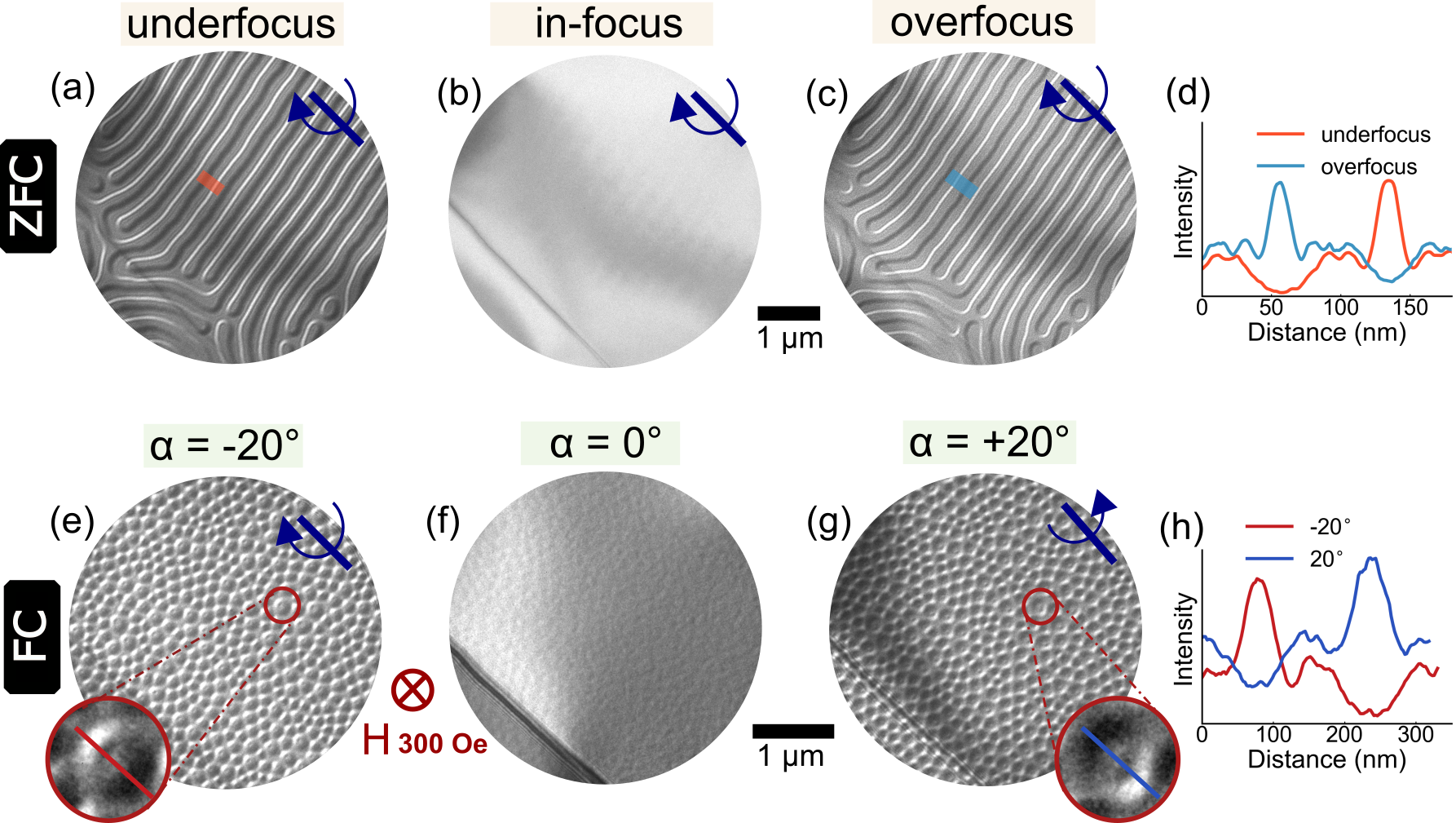}
\caption{ Lorentz TEM images of N\'eel-type stripe domain walls under (a) underfocus, (b) in-focus and (c) overfocus conditions at 100 K after zero-field cooling (ZFC). Sample tilt angle $\alpha$ is $-20\degree$. (d) Average line profile across the orange and light blue bold lines in (a) and (b). (e-g) Lorentz TEM images of N\'eel-type skyrmions with different sample tilt, $\alpha$, after field cooling (FC) in a 300 Oe out-of-plane field. (h) Line profiles across the skyrmions shown in insets of (e) and (g).}
\label{fig:Neel_DWs}
\end{figure}

The \ce{Fe_3GeTe_2} single crystals were grown by a chemical vapor transport method using \ce{I_2} as the transport agent. A stoichiometric mixture of Fe, Ge, and Te powders, together with the transport agent, were sealed in an evacuated quartz tube and heated in a two-zone tube furnace with a temperature gradient from 700 to 600 \degree C for 1 week. Millimeter-size single crystals were obtained. The compositions of the crystals were determined by energy-dispersive x-ray spectroscopy. Thin flakes of $\FGT$ were mechanically exfoliated from the bulk crystal in an ambient atmosphere. The exfoliated flakes were picked up by a polydimethylsiloxane (PDMS) stamp and placed on the electron-transparent window of a silicon nitride TEM grid. The flakes were successfully transferred to the silicon nitride membrane by heating the sample to around \SI{80}{\celsius}, at which the PDMS becomes less sticky. The transferred flakes were then coated with a 3 nm thick aluminum layer using DC sputtering to prevent further oxidation. The thickness of the \ce{Fe_3GeTe_2} flake discussed in this paper was determined to be approximately 160 nm by using atomic force microscopy. We characterized the magnetic domains using the out-of-focus Fresnel Lorentz mode\cite{portier1999electron, tanase2009situ} in a FEI Tecnai F20ST TEM instrument operating at 200 kV. The cryo-TEM experiment was carried out using a Gatan liquid nitrogen cryo-holder, which can reach a minimum temperature of 100 K. 

Figure ~\ref{fig:Neel_DWs}(a-c) show a through-focus series of LTEM images of \ce{Fe_3GeTe_2} with the sample tilted to $\alpha = $ $-20\degree$. The images were recorded at 100 K after the sample was zero-field cooled (ZFC) from room temperature, i.e. cooling in the absence of any external magnetic field. The Curie temperature of the \ce{Fe_3GeTe_2} flake is about 200 K based on our $in$ $situ$ LTEM measurements. Alternate bright and dark contrast appears at the location of the domain walls in the out-of-focus images (Figs.~\ref{fig:Neel_DWs}(a) and (c)), which disappears under the in-focus condition (Fig.~\ref{fig:Neel_DWs}(b)). This suggests that the contrast originates from magnetic domains rather than from the microstructure. The absolute value of defocus is 2 mm for both the under- and over-focus images. The locations of the bright and dark contrast reverses from underfocus to overfocus, as shown in the line profiles in Figs.~\ref{fig:Neel_DWs}(d). 

Fig.~\ref{fig:Neel_DWs}(e) shows the formation of densely-packed skyrmions at 100 K when the sample was field cooled (FC) under an externally-applied 300 Oe magnetic field aligned with electron beam (see the schematic of Fig.~\ref{fig:deform}(a)). When the sample is tilted to $\alpha = $ $0\degree$ as shown in Fig.~\ref{fig:Neel_DWs}(f) the magnetic contrast almost disappears for the same defocus length of 2 mm. The magnetic domain contrast is observed again with opposite contrast when the sample is tilted to $\alpha = $ $+20\degree$, as shown in Figs.~\ref{fig:Neel_DWs}(g) and the line plot in Figs.~\ref{fig:Neel_DWs}(h). It should be noted that the stripe domain walls also show the same behavior, which is not shown here. This suggests that the magnetization in the domains is oriented in a direction parallel to the electron beam i.e. perpendicular to the plane of the sample, and that the domain walls exhibit a N\'eel-type structure (see insets of Figs.~\ref{fig:exp_vs_sim}(e) and (h)), where the in-plane spins point in a direction perpendicular to the plane of the domain wall. In LTEM, the electron beam is sensitive only to the component of magnetic induction that is perpendicular to the electron propagation direction. Therefore, at zero sample tilt angle, there is no contribution to the image contrast from the magnetic domains. Domain walls with N\'eel-type character result in diverging or converging behavior of the electrons, and thus give rise to a net zero phase shift of the electrons\cite{jiang2019quantifying, benitez2015magnetic}. Hence no domain wall contrast is observed.  
However, when the sample is tilted, a component of the out-of-plane magnetization in the domains is projected to lie perpendicular to the direction of the electron beam, giving magnetic contrast that can be observed in the out-of-focus LTEM images. For N\'eel-type skyrmion domain, a similar effect gives rise to bright and dark semi-circle contrast at each skyrmion location, as shown in the inset of Figs.~\ref{fig:Neel_DWs}(e). To further understand and correlate the projected magnetic induction maps with the magnetization of the sample, we performed micromagnetic simulations along with LTEM image simulations to compare with the experimental results, as presented in the next section.


\subsection*{Retrieval of magnetization configurations }\label{subsec:exp_sim}

\begin{figure}[h]%
\centering
\includegraphics[width=0.9\textwidth]{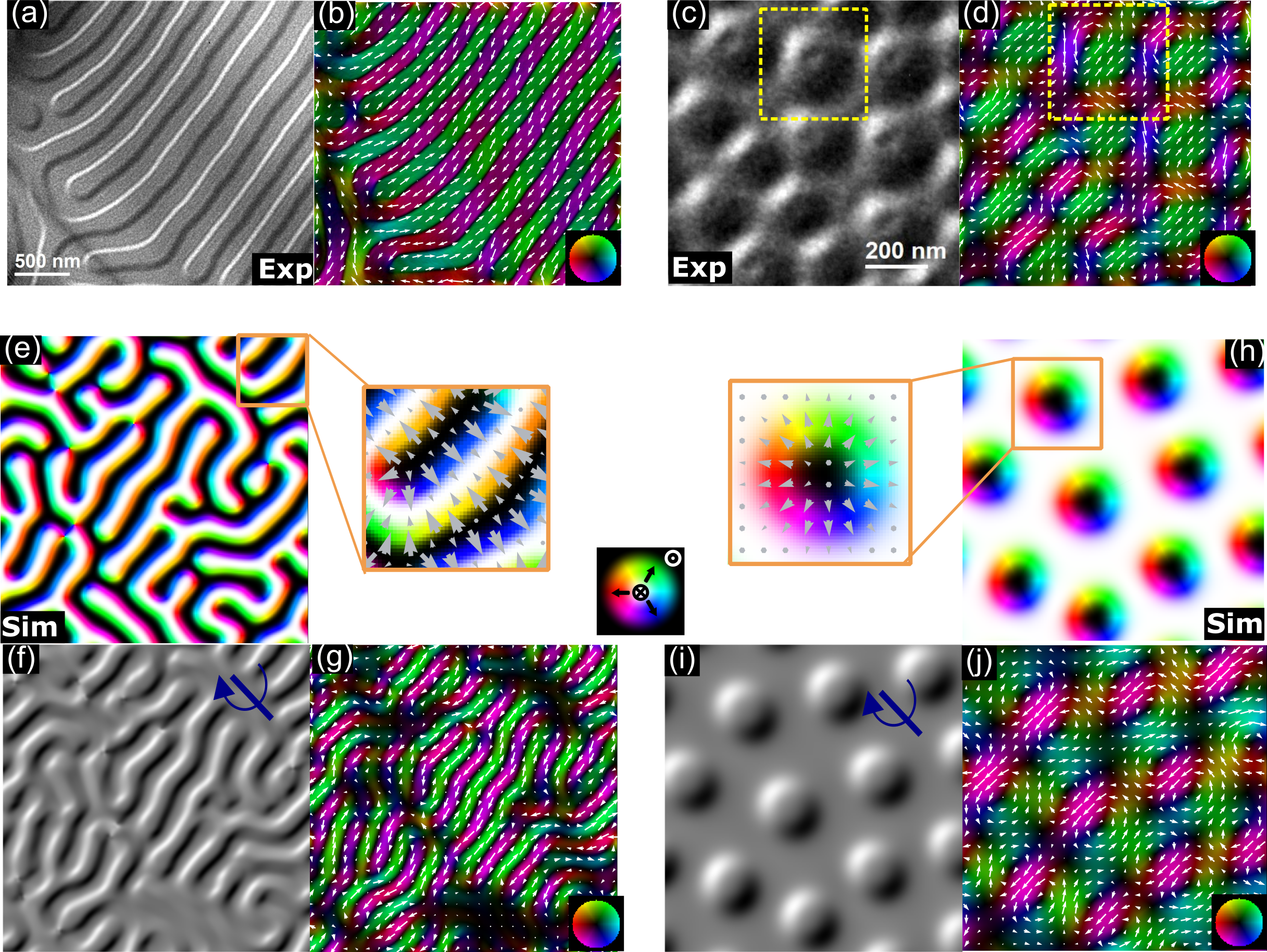}
\caption{(a, c) Experimental Lorentz TEM images and (b, d) reconstructed in-plane magnetic induction maps of N\'eel stripe and skyrmion domains. (e, h) Simulated magnetic spin textures with magnified insets showing magnetization orientation of color indicated by arrows, (f, i) simulated Lorentz image contrast and (g, j) reconstructed magnetic induction maps for N\'eel-type stripe and skyrmions domains. Colorwheels represent magnetization direction.}
\label{fig:exp_vs_sim}
\end{figure}

To better interpret the N\'eel-type magnetic domain contrast seen in our LTEM images, we retrieved the magnetic phase shift, which is related to the integrated magnetic induction, from a through-focus series of Fresnel images  using the open-source 'PyLorentz' code based on the transport-of-intensity equation\cite{mccray2021understanding}. Figs.~\ref{fig:exp_vs_sim}(a-d) are experimental LTEM images and the corresponding reconstructed magnetic induction maps for stripe domains and skyrmions respectively. Figs.~\ref{fig:exp_vs_sim}(a) and (b) show explicitly that the bright and dark contrast stripes seen in the images correspond to the positions of the domain walls separating adjacent domains with opposite magnetization orientations. As explained previously, the in-plane magnetic moment seen in  Fig.~\ref{fig:exp_vs_sim}(b) is introduced by the projection of the out-of-plane magnetization of \ce{Fe_3GeTe_2} due to sample being tilted \cite{ding2019observation, yang2020creation}. We observe similar contrast for the N\'eel skyrmions seen in  Figs.~\ref{fig:exp_vs_sim}(d), with a uniform magnetic induction inside each skyrmion (green) and the projected stray field outside the skyrmions (pink). Contrast for a single skyrmion is marked by the yellow dashed frame in Figs.~\ref{fig:exp_vs_sim}(c) and (d).

The sample tilt leads to interpretation of the projected magnetic induction maps being more complex. In addition, the reconstructed LTEM images represent the total projected magnetic induction, rather than just the magnetization in the sample.\cite{mccray2021understanding} Further analysis of the magnetic domain structure therefore requires us to calculate the LTEM contrast obtained from micromagnetic simulations for comparision with experiment. We performed micromagnetic simulations using the MuMax package \cite{vansteenkiste2014design}, to generate N\'eel-type magnetization configurations using the parameters for $\FGT$. The simulation grid size for the data shown in Fig.~\ref{fig:exp_vs_sim}(e) is $1100 \times 1100$ nm$^2$ with a 5 nm cell size, and the grid size shown in Fig.~\ref{fig:exp_vs_sim}(e) is $400 \times 400$ nm$^2$ with a 2 nm cell size. The larger grid size for the stripe domain simulation is needed to reproduce the extended domains, and a smaller cell size is used for the skyrmions in order to simulate their spin configuration with higher spatial resolution. The saturation magnetization, exchange constant and first-order uniaxial anistropy constant were set to be $3.7\times 10^5$ A/m, $1.4\times 10^{11}$ J/m and $1.1 \times 10^6$ J/m$^3$, respectively. We set the value of the interfacial Dzyaloshinskii-Moriya interaction (DMI) strength as 1.6 mJ/m$^2$. The saturation magnetization and uniaxial anistropy constant are obtained from experiment (not shown here) and the other simulation material parameters are optimized in order to obtain similar magnetization textures to those seen experimentally.

The simulated stripe domains and skyrmions shown in Figs.~\ref{fig:exp_vs_sim}(e) and (h) are formed by following a relaxation process from a random initial state to a final equilibrium state without, or with, the presence of an external magnetic field. The two magnified insets show the magnetization configurations of N\'eel-type domain walls, wherein the spins lie in-plane and point towards one of the adjacent out-of-plane domains. We simulate the LTEM images from the output of the micromagnetic simulations using the linear superposition method\cite{mccray2021understanding}, for a sample tilt direction as indicated in the inset of Figs.~\ref{fig:exp_vs_sim}(f) and (i), which is same as that in the LTEM measurement. It can be seen that the simulated results show a good agreement with the experimental data. In Fig.~\ref{fig:exp_vs_sim}(f) the magnetic stripe domains that are perpendicular to the tilt axis show strong bright and dark contrast, whereas those that are parallel to the tilt axis show very weak contrast.  The simulation of the skyrmions is also very consistent with experiment: bright and dark contrast in the out-of-focus LTEM image (Fig.~\ref{fig:exp_vs_sim}(i)) arising from the convergence and divergence of electron beams deflected by projected components of skyrmions inside (pink) and outside (green) stray magnetic fields, as shown in Figs.~\ref{fig:exp_vs_sim}(d) and (j). It is also worthy to notice that the bright and dark semi-circle contrasts are aligned with the tilting direction (see Figs.~\ref{fig:exp_vs_sim}(c) and (i)). By comparing with simulations, we can better understand the experimentally-observed magnetic spin textures of the tilted N\'eel domains.


\subsection*{Deformed Néel skyrmion contrast}\label{subsec:def_const}

\begin{figure}[h]%
\centering
\includegraphics[width=0.8\textwidth]{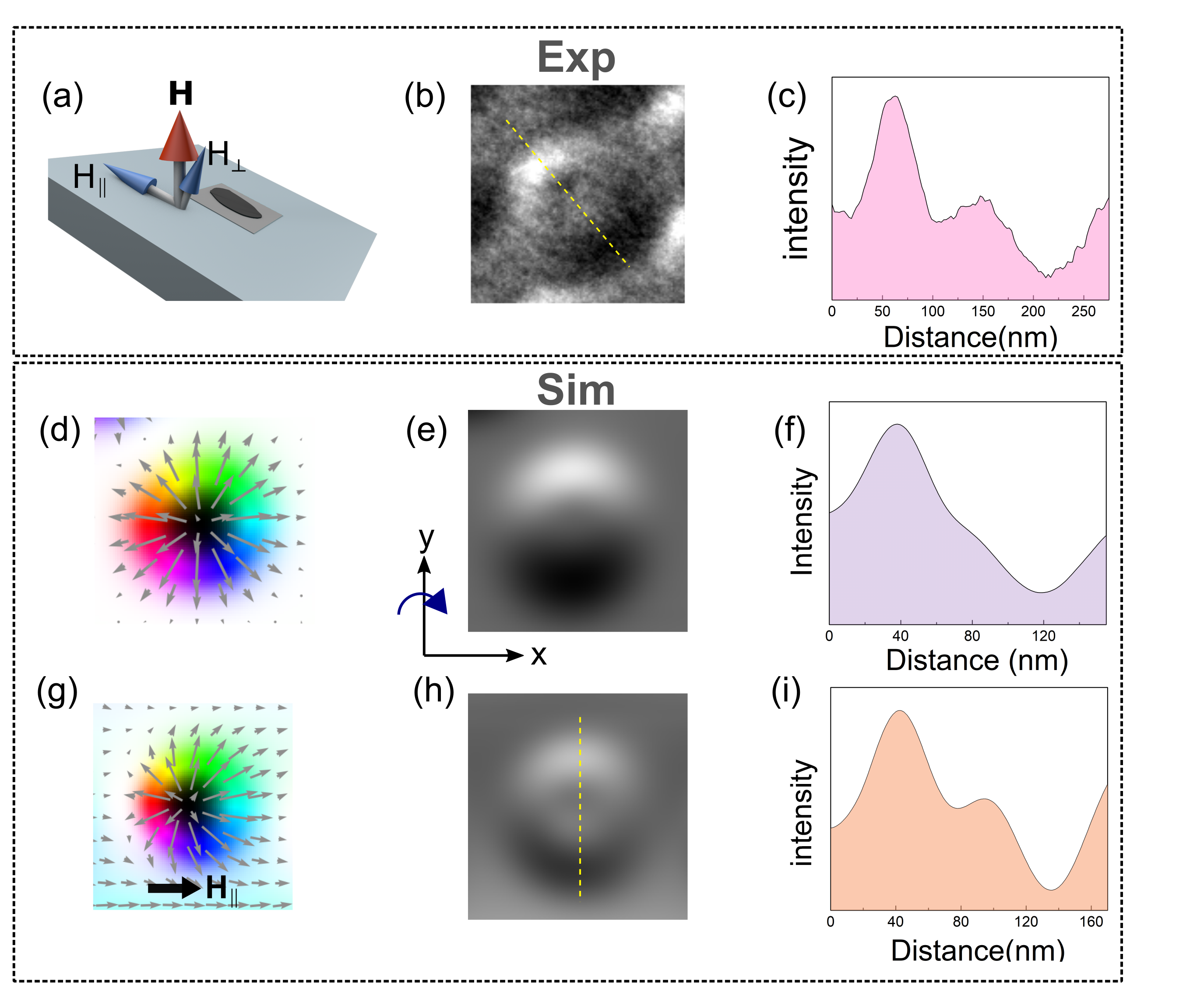}
\caption{(a) Schematic shows the effective direction of the of in-plane ($H_{\parallel}$) and out-of-plane ($H_{\perp}$) magnetic fields introduced while the \ce{Fe_3GeTe_2} sample is tilted during the TEM experiments. (b) Experimental Lorentz TEM image of a deformed N\'eel skyrmion and (c) the intensity profile across the yellow dashed line in (b). (d and g) Simulated magnetization configurations, (e and h) Lorentz TEM image contrast and (f and i) line profiles across the simulated images for a standard N\'eel skyrmion (d-f) and a deformed N\'eel skyrmion (g-i).}
\label{fig:deform}
\end{figure}

Interestingly, we observe additional contrast in the center of some of the larger-size skyrmions, leading to a bright-dark-bright-dark contrast structure, as shown in Figure~\ref{fig:deform} (b). This additional contrast is relatively weak, as shown by the line plot Fig.~\ref{fig:deform}(c) taken along the dashed line in Fig.~\ref{fig:deform}(b). One hypothesis for the origin of this contrast is that it arises from a so-called N\'eel skyrmionium, which is composed of two twisting vortex cores with zero topological charge\cite{zhang2018real, kolesnikov2018skyrmionium, zhang2020deformation}. The bright-dark-bright-dark pattern is generated from the projected magnetization of two cores with opposite out-of-plane magnetic moments. However, when we simulated LTEM images from a simulated skyrmionium spin texture, we see that the intensity of the inner dark and bright contrast regions is as strong as that from the outer ones, which disagrees with our experimental result. The other possibility is a deformation of the skyrmion contrast because of an in-plane component of applied perpendicular magnetic field introduced due to sample tilt (see Fig.~\ref{fig:deform}(a)). A similar magnetic texture was previously observed in a [Pt/Co/Ta]$_{20}$ multilayer sample by Zhang \textit{et al}\cite{zhang2020deformation}.  In our experimental data, we also only see the additional contrast inside the skyrmion upon field cooling when the sample is tilted. 

In order to verify this hypothesis, we simulated the spin texture of a skyrmion in the presence of a magnetic field with both out-of-plane ($H_{\perp} = 1500$ Oe) and in-plane ($H_{\parallel} = 200$ Oe) components. In contrast to the spin texture of a perfect N\'eel skyrmion shown in Fig.~\ref{fig:deform}(d), Fig.~\ref{fig:deform}(g) shows that deformed configuration occurs because the spins attempt to align with in-plane magnetic field component, leading to an onion-like magnetization configuration. The small magnetic moment component in the domain walls that is aligned with the external in-plane field component contributes to the additional weak dark and bright contrast in the skyrmion center. A simulated LTEM image (see Fig.~\ref{fig:deform}(h))  of Fig.~\ref{fig:deform}(g), titled around the $y$-axis by $-20\degree$, shows the identical bright-dark-bright-dark contrast to that seen experimentally, with the middle dark and bright contrast regions having lower intensity as displayed in the line profile in Fig.~\ref{fig:deform}(i). However, we do not see the deforming bright-dark-bright-dark  contrast in the smaller skyrmions. This may result from the overlapping of the weak contrast of the deformed skyrmion with the strong contrast at the outer edge of the skyrmion. It is also worthy to note that in the experiment the introduced $H_{\parallel} = 103 $ Oe as a 300 Oe magnetic field was applied during FC and the sample was tilted by $\alpha=-20\degree$. The ratio of two magnetic field components ($H_{\parallel}/H_{\perp}$) are different from that used in the simulation, as a result of fact that micromagnetic simulation cannot consider all impacts on the skyrmion formation in the experiment, for example nonuniform strains and thickness variations in the exfoliated $\FGT$. We can therefore conclude that the FC process combined with the sample tilt is sufficient to deform the spin textures associated with the skyrmions.


\subsection*{$In$-$situ$ magnetizing experiments}

So far we have shown that ZFC and FC of \ce{Fe_3GeTe_2} flake across the Curie temperature results in two distinct magnetic structures, namely N\'eel stripe domains and N\'eel skyrmions, respectively. Next, we explore the response of these different magnetic domains to an external magnetic field applied $in$ $situ$ at a fixed low temperature in order to understand their stability and domain behavior. 

\subsubsection*{Field-dependent behavior of stripe domains}

\begin{figure}[ht]%
\centering
\includegraphics[width=\textwidth]{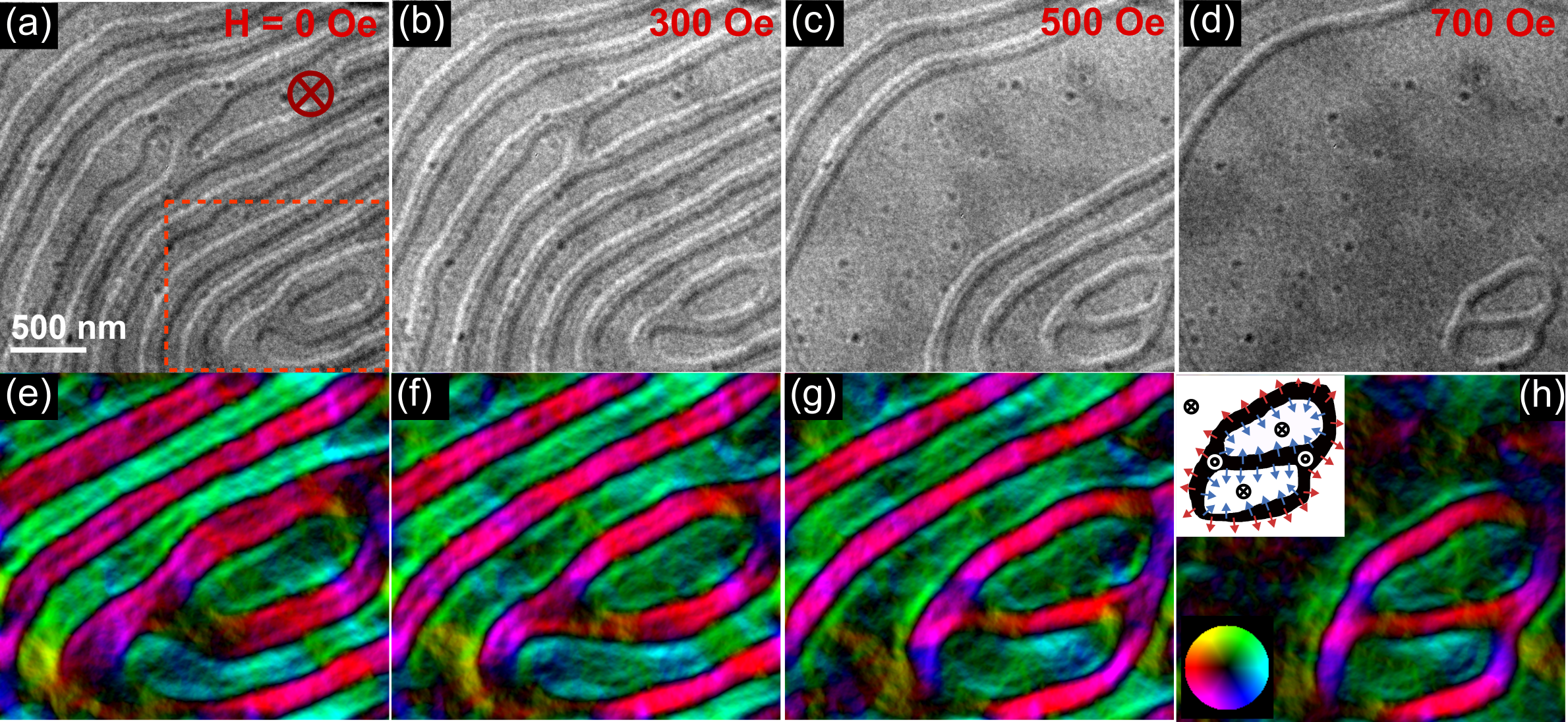}
\caption{ Magnetic-field-dependent evolution of N\'eel stripe domains. (a-d) Experimental Lorentz TEM images and (e-h) reconstructed projected component of magnetic induction. The red dashed frame in (a) highlights the sample region for the reconstructed magnetization configuration maps. The inset of panel (h) presents the schematic of one possible spin texture for domain wall of two adjacent closed-loop domains where arrows represent magnetization orientations. }
\label{fig:B_maze}
\end{figure}

Figures~\ref{fig:B_maze}(a-h) show out-of-focus LTEM images and the corresponding reconstructed magnetic induction maps of the bottom-right region (marked by the red dash frame in Figs.~\ref{fig:B_maze}(a)) when the sample undergoes a ZFC process. The resulting domain structure consists of stripe domains. This particular exfoliated flake of $\FGT$ was measured to be 60 nm thick and was coated with a 3 nm Pt capping layer. There was no difference in the behavior noted due to capping with Pt versus capping with Al, which suggests that the capping layer does not strongly affect the interfacial DMI and resulting spin textures of magnetic domains in this material. Figs.~\ref{fig:B_maze}(a) and (e) show the initial N\'eel stripe domains at 100 K before application of the external magnetic field. As the applied magnetic field perpendicular to the sample is increased, we observe that the width of the magnetic domains decreases for those domains in which the magnetization orientation is anti-aligned with the direction of the applied field, as shown in Figs.~\ref{fig:B_maze}(b) and (f). From Fig.~\ref{fig:B_maze}(b) to (c) it can be seen that the density of stripe domain decreases as the anti-aligned domains (pink region) shrink and disappear eventually under a higher magnetic field. This suggests that magnetic field can control the density and size of stripe domains.  We also notice that a complex closed-loop magnetic domain configuration at the bottom right corner (in the red dash frame of Figs.~\ref{fig:B_maze}(a)) is significantly stable up to magnetic fields of 700 Oe. At 500 Oe, the second closed loop forms adjacent to the first one (Figs.~\ref{fig:B_maze}(d)). The additional stability of this domain can be explained due to the total topological charge of the enclosed magnetic domain which prevents it from collapsing\cite{braun2012topological,rybakov2019chiral}. This behavior is also similar to the stability of adjacent $360\degree$ domain walls\cite{cubukcu2018electrical}. One of possible spin textures of two adjacent $360\degree$ domain walls is schematically presented in inset of Fig.~\ref{fig:B_maze}. The \ce{Fe_3GeTe_2} flake was fully polarized at an applied field strength of 1000 Oe.

Here we can see that if the sample has been ZFC to 100 K, a uniform skyrmion phase cannot be formed even if perpendicular magnetic field is then applied at this temperature. We observe the formation of only isolated closed-loop magnetic domains as the stripe domains shrink in response to the applied field, but they are unable to change forming a skyrmion lattice, as that would require overcoming a large energy barrier associated with deformation of domain walls. 
In order to achieve a uniform skyrmion phase, we need to apply a magnetic field during the cooling process across the phase transition from the paramagnetic to the ferromagnetic state. This can be explained on the basis that as the material is undergoing the phase transition close to the Curie temperature, the Zeeman energy is the dominant energy term as compared to other isotropic and anisotropic magnetic energy terms (exchange, magnetocrystalline anisotropy, and DMI). 
Thus in the absence of external field, the magnetization would form an extended spin spiral resulting in stripe domains whereas the presence of external field prevents such extended domains from forming, creating localized spin textures of skyrmions to minimize the total energy of the system.

\subsubsection*{Field-driven behavior of skyrmions}
\begin{figure}[h]%
\centering
\includegraphics[width=1\textwidth]{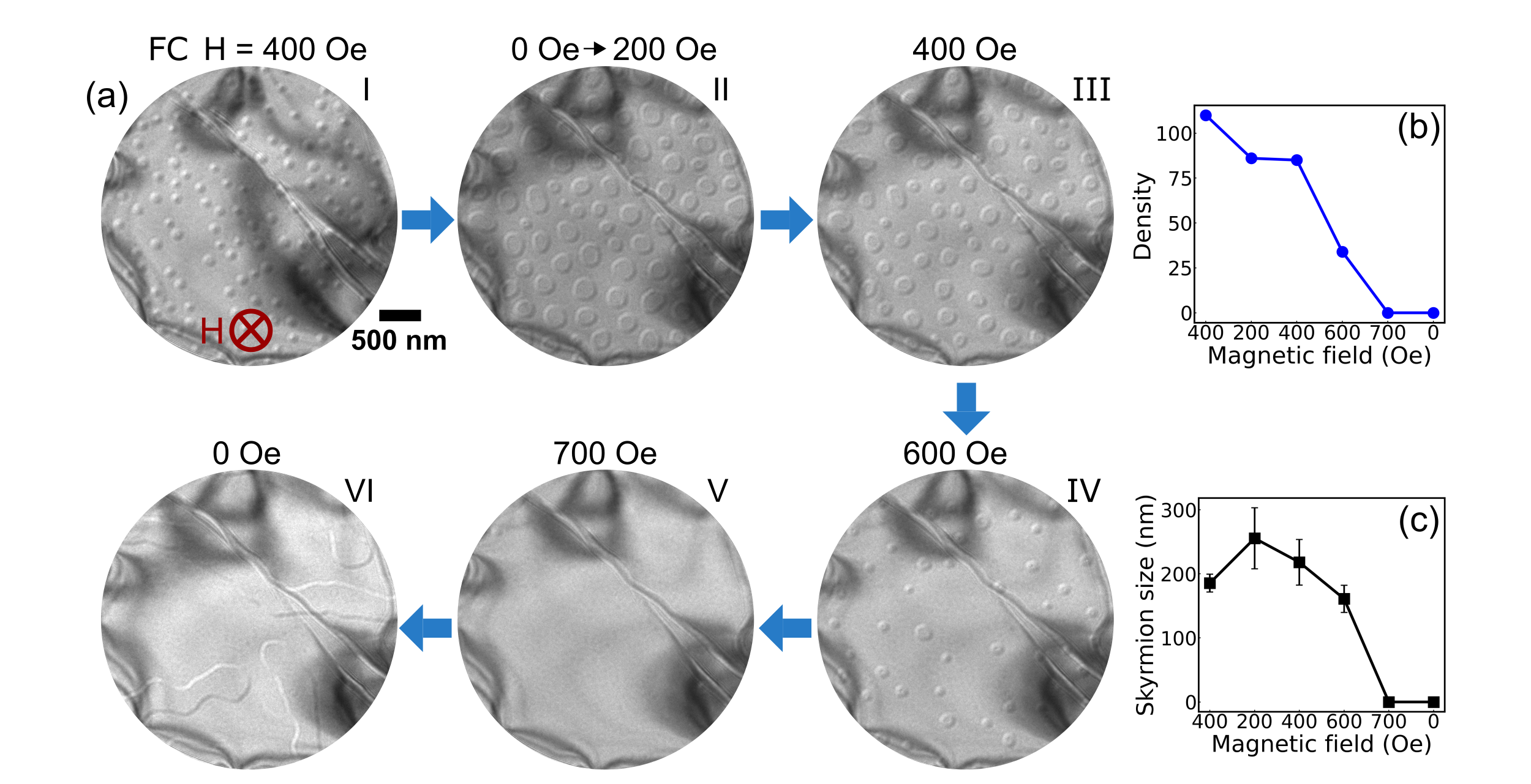}
\caption{(a) Response of N\'eel skyrmions in $\FGT$ to removal and then increase of an $in$ $situ$ perpendicular applied magnetic field at 100 K. (b) Density and (c) average size of the skyrmions as a function of magnetic field. }
\label{fig:B_skyrm}
\end{figure}

We also examine how N\'eel skyrmions behave under an external magnetic field in the same \ce{Fe_3GeTe_2} sample. Figure~\ref{fig:B_skyrm}(a)I shows loosely-packed N\'eel skyrmions at a temperature of 100 K, generated after FC in a 400 Oe field. The thickness of this flake ($\sim$60 nm) is smaller than that of the flakes shown in Figs. 1 and 2 ($\sim$160 nm), and so the size and density of the skyrmions are also lower. We then first relax the skyrmion phase in \ce{Fe_3GeTe_2} by removing the external magnetic field, followed by applying a 200 Oe field aligned with electron beam in TEM. The average size of skyrmions increases from 185 nm to 255 nm (with a large standard deviation)  while the skyrmion density drops as a result of neighboring skyrmions merging, as shown in Fig.~\ref{fig:B_skyrm}(a) II. In addition, the skyrmion size is much less uniform. Increasing the magnetic field further (Fig.~\ref{fig:B_skyrm}(a) III-IV) leads to a decrease in both size and density of the skyrmions. The sample is observed to be fully polarized at an applied field of 700 Oe, which is in contrast to the ZFC case, where an external field of 1000 Oe is needed to achieve the fully polarized state. Figs.~\ref{fig:B_skyrm}(b) and (c) show plots of the change in skyrmion density and size respectively as a function of the applied magnetic field. Upon removal of the applied magnetic field after saturation, we see a return to a phase with broad magnetic domains rather than to a uniform skyrmion phase. This suggests that the skyrmion phase is a metastable state that is formed during the FC process. This phase is stabilized, as explained previously, due to the dominance of the Zeeman energy, which breaks the symmetry of the system and results in the formation of skyrmions. At low temperatures, the application of a sufficiently strong external field causes the Zeeman energy to become dominant energy term again, which is able to break the topological protection of the skyrmions\cite{tokura2020magnetic} and results in a saturated magnetization state. When the external field is then removed, the sample returns to the stripe domain phase similar to the ZFC conditions, which is energetically more favorable as the equilibrium state.

\FloatBarrier
\section*{CONCLUSIONS}

In this work we examine systematically the role of magnetic field on the formation and manipulation of magnetic domain typologies in vdW $\FGT$. We observe N\'eel-type stripe domains and skyrmions in \ce{Fe_3GeTe_2} after following a ZFC and FC protocols respectively. A combination of micromagnetic simulation and quantitative analyses of LTEM images allows us to understand the magnetic phase contrast of tilted N\'eel-type magnetic domains in LTEM images. The sample tilting also leads to a small in-plane magnetic field introduced during the FC process, which deforms the spin texture of skyrmions resulting in the additional contrast observed in the center of larger skyrmions. The behavior of the magnetization configurations for the N\'eel stripe domains and skyrmions in response to an external magnetic field applied at low temperature are different from those observed during the cooling process. The bias magnetic field controls the size and density of the existing magnetic domains but cannot change stripe domains to skyrmions or vice versa due to the energy barrier of domain walls. A change in domain topology at low temperature, i.e. from skyrmion to stripe domain, only occurs when the skyrmion magnetization configuration is first completely annihilated by an applied magnetic field. We also showed that the skyrmion phase in $\FGT$ is metastable and can only be accessed during the FC process. Overall, our findings demonstrate that Zeeman energy term leads to different magnetic domain behaviors depending on whether the magnetic field is applied at above or below Curie temperature.  Understanding how to control the field-dependent behavior of the topological magnetic domains discussed here paves the way to the practical applications of 2D vdW materials for spintronics.

\section*{ACKNOWLEDGMENTS}

This work was funded by the US Department of Energy,
Office of Science, Office of Basic Energy Sciences, Materials Science and Engineering Division. Use of the Center for Nanoscale Materials, an Office of Science user facility, was supported by the U.S. Department of Energy, Office of Science, Office of Basic Energy Sciences, under Contract No. DE-AC02-06CH11357. Single crystal growth is supported by the U.S. Department of Energy, Office of Science, Basic Energy Sciences program under Grant No. DE-SC0022006.

The submitted manuscript has been created by UChicago Argonne, LLC, Operator of Argonne National Laboratory (“Argonne”). Argonne, a U.S. Department of Energy Office of Science laboratory, is operated under Contract No. DE-AC02-06CH11357. The U.S. Government retains for itself, and others acting on its behalf, a paid-up nonexclusive, irrevocable worldwide license in said article to reproduce, prepare derivative works, distribute copies to the public, and perform publicly and display publicly, by or on behalf of the Government. The Department of Energy will provide public access to these results of federally sponsored research in accordance with the DOE Public Access Plan. http://energy.gov/downloads/doe-public-access-plan

\section*{CONFLICT OF INTEREST}
The authors declare that they have no conflict of interest.

\bibliography{ref}

\end{document}